\documentclass[aps,pre,preprint,groupedaddress]{revtex4}

\usepackage[dvips]{graphicx}
     
\def\bd{\begin{displaymath}}
\def\be{\begin{equation}}
\def\ed{\end{displaymath}}
\def\ee{\end{equation}}

\begin{document}

\title{\bf Electromigration-Induced Propagation 
of Nonlinear Surface Waves} 
\author{R. Mark Bradley}
\email[]{bradley@lamar.colostate.edu}
\affiliation{
Department of Physics,
Colorado State University,
Fort Collins, CO 80523 USA}

\date{\today}

\begin{abstract}

Due to the effects of surface electromigration,
waves can propagate over the free surface of a current-carrying 
metallic or semiconducting film of thickness $h_0$.   
In this paper, waves of finite
amplitude, and slow modulations of these waves, are studied.  Periodic wave trains
of finite amplitude are found, as well as their dispersion relation.
If the film material is isotropic, 
a wave train with wavelength $\lambda$ is unstable if 
$\lambda/h_0 < 3.9027\ldots$,
and is otherwise marginally stable.  The equation of motion for slow modulations
of a finite amplitude, periodic wave train is shown to be 
the nonlinear Schr\"odinger equation.  As a result, envelope solitons can 
travel over the film's surface.

\end{abstract}


\maketitle

{\bf I. Introduction}

Waves propagating over the free surface of a layer of water
subject to gravity have been studied extensively for well over a 
century.  In his ground breaking paper of 1847, Stokes showed 
that steady periodic wave trains of finite amplitude are possible,
and that their dispersion relation depends upon their amplitude \cite{Stokes}.
In these wave trains, the effects of dispersion and nonlinearity
balance, yielding a steady wave form.

In the 1960's, it was discovered that if the water is of finite
depth $h_0$ and the wavelength $\lambda=2\pi/k$ is sufficiently short,
the wave train is unstable.  Benjamin and Feir \cite{BF} 
showed that
the Stokes wave train is unstable to small modulational or
side band perturbations if $k h_0 > 1.363\ldots$,
and this been confirmed experimentally \cite{Benjamin,Yuen,Lake,Shemer}.

Benjamin and Feir's analysis holds only at early times, just after 
the instability has begun to develop.  Zakharov \cite{Zakharov} and
later Hasimoto and Ono \cite{HO} derived an equation of motion for 
the envelope of a packet of finite-amplitude 
gravity waves which remains valid long after the onset of the instability.
Their equation --- the nonlinear
Schr\"odinger equation --- admits envelope soliton solutions
that are intimately related to the ultimate state of the fluid surface.

Although it is not yet widely appreciated, there are nontrivial 
{\it electrical} free boundary problems. 
When an electrical current passes through a piece of solid metal
or semiconductor, 
collisions between the conduction electrons and the atoms at the
surface lead to drift of these atoms.  This phenomenon, which is 
known as surface electromigration (SEM), can cause the surface of
a solid to move and deform 
\cite{Latyshev,Stoyanov,Krug_and_Dobbs,Kraftapl,Suojap,Mohanjap,Krug1,Krug2,Gengor1,Gengor2,Gengor3,Mohanepl,Mohanpre}.  
The free surface of a conducting film moves in response
to the electrical current flowing through the bulk of the film, in much
the same way that flow in the bulk of a fluid affects the motion of 
its surface.  However, the analogy is not perfect --- the 
boundary conditions are very different in the two problems.
In addition, metals and semiconductors are anisotropic materials,
and this distinguishes them from fluids as well.

Waves can propagate over the free surface of a current-carrying 
film deposited on an insulating substrate, at least in the
limit of small amplitude.  These waves, whose propagation is 
induced by surface electromigration, 
are somewhat analogous to gravity waves, and will be
called \lq\lq SEM waves."  The linear dispersion relation for these
waves has been known for some time \cite{Krug1}, and will be discussed 
briefly in Sec. III.  

It is natural to ask to what extent the 
nonlinear behavior of gravity waves is paralleled by that 
of SEM waves.  To address
this question, in this paper I will study SEM waves of finite
amplitude, and slow modulations of these waves.  Periodic wave trains
of finite amplitude (the analog of the Stokes wave) are found, as well
as their dispersion relation.
If the film material is isotropic, the wave trains are unstable for $kh_0 > 1.609927\ldots$
and are otherwise marginally stable.  The equation of motion for slow modulations of the finite amplitude, periodic wave train is shown to be 
the nonlinear Schr\"odinger equation. 
  
It is worth mentioning that SEM is not just of academic interest: 
In modern integrated circuits, the current-carrying metal lines or \lq\lq interconnects" carry very high current densities, 
and electrical failure can occur due to the effects of SEM.
In particular, SEM can cause a small perturbation at the edge of a
metal strip to become a slit-shaped void  
which propagates across the line until its tip contacts 
the opposite side of the strip; electrical failure is then complete
\cite{Mohanepl,Mohanpre,Rose,Sanchezapl}. 

{\bf II. Equations of Motion}

Consider a single-crystal metallic or semiconducting film 
of uniform thickness 
$h_0$ deposited on the plane surface of
an insulating substrate.  
We take the $z$ axis to be normal to the film-vacuum interface 
and locate the origin $O$ in this plane.  (We will
occasionally find it convenient to use another set of Cartesian coordinates
$(x', z')$ with $x'=x$, $z'=z+h_0$ and origin $O'$ at the substrate
surface.)  The film's surface
$z=0$ will be assumed to be a low index crystal plane. 
A constant electrical current flows
through the film in the $x$ direction, and the electric field within 
the film is $\vec E_0 = E_0\hat x$.

\begin{figure}
\includegraphics[scale=1]{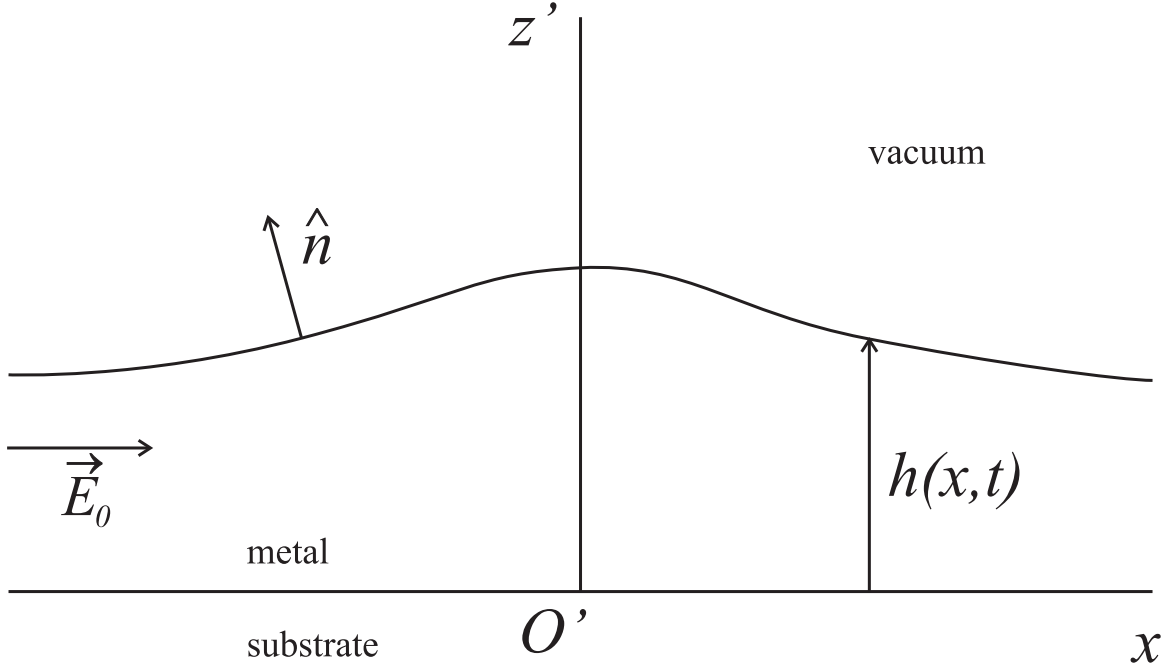}%
\caption{\label{Fig. 1} The current-carrying metal thin film.  The height of the free surface 
above the substrate, $h$, depends only on $x$ and $t$.  The outward-pointing
unit normal to the free surface is $\hat n$, and the electric field far from the 
perturbation is $\vec E_0$.
}
\end{figure}

Now suppose that the upper surface of the 
film is perturbed. Let the 
outward-pointing unit normal to this surface be $\hat n$.  For simplicity,
we shall restrict our attention to perturbations whose form does not depend
upon $y$, so that the height of the film's surface above the substrate $h=h_0+\zeta$
depends only on $x$ and $t$ (Fig. \ref{Fig. 1}).  The upper film surface will evolve in the 
course of time due to the effects of SEM and surface self-diffusion.  We 
assume that the current flowing through the film is held fixed.

Clearly, the problem is two-dimensional (2D), 
and the dependence of all quantities
on $y$ will therefore be suppressed.  
The electrical potential $\Phi=\Phi(x,z,t)$ satisfies
the 2D Laplace equation
\be
\nabla^2\Phi=0
\label{orig_Laplace}
\ee
and is subject to the boundary condition  
$\hat n\cdot\vec\nabla\Phi = 0$ on the upper surface and 
$\hat z\cdot\vec\nabla\Phi = 0$ on the lower.
More explicitly, we have
\be
\Phi_z(x,\zeta,t) =\zeta_x\Phi_x(x,\zeta,t)
\label{orig_topbc}
\ee
and
\be
\Phi_z(x,-h_0,t)=0,
\label{orig_bottombc}
\ee
where $f_x\equiv\partial f/\partial x$ and so forth.  
If the initial perturbation
is localized, we also have
\be
\zeta\to 0
\label{orig_loc_h}
\ee
for $x\to\pm\infty$ and, furthermore,
\be
\vec\nabla\Phi(x,z,t)\to -\vec E_0
\label{orig_loc_phi}
\ee
for $x\to\pm\infty$ and $-h_0\le z \le 0$.

We assume that the atomic mobility is negligible at the 
film-insulator interface, so that the form of that interface remains
planar for all time.  Further, in the interest of simplicity, we assume
that the applied current is high enough that the effects of SEM are 
much more important than those of capillarity. The surface atomic
current $\vec J_a$ is then proportional to the electrical current at the surface.
Explicitly, $\vec J_a = -\nu Mq\vec\nabla\Phi$, where the $\nu$
is the areal density of the mobile surface atoms, $q$ is their effective 
charge, and $M$, the adatom mobility, in general depends on the surface slope
$\zeta_x$.  When there is a net influx of atoms into a small surface element, it will move.  
The normal velocity of the film's free surface is 
$v_n = q\nu\Omega \partial_s(M\partial_s \Phi)$,
where $\Omega$ is the atomic volume and $s$ is the arc length along the surface. 
Therefore 
\be
\zeta_t=q\nu\Omega\partial_x\left [ {{M(\zeta_x)}\over{\sqrt{1+\zeta_x^2}}}
	\partial_x\Phi(x, \zeta, t)\right ],
\label{orig_eom}
\ee
where $\partial_x$ is the total derivative with respect to $x$.  
(Since $\zeta$ depends on $x$, we have 
$\partial_x\Phi(x,\zeta,t) = \Phi_x(x,\zeta,t)+\zeta_x\Phi_z(x,\zeta,t)$, for example.)
Together, Eqs. (\ref{orig_Laplace}) - (\ref{orig_eom}) completely describe the nonlinear dynamics of the film surface.

{\bf III. Finite Amplitude Periodic Wave Trains}

We wish to study the propagation of a disturbance whose 
amplitude is small. 
To this end, we put $\Phi = -E_0 x +\phi$ and presume
that $\zeta$ and $\phi$ are of order $\epsilon$.

We shall suppose that the crystal structure is invariant 
under the reflection $x\to -x$, so that $M$ is an even function of
the slope $\zeta_x$.  As a result,  
$M(\zeta_x) = M_0(1+c\zeta_x^2)+O(\epsilon^4)$,
where $M_0 \equiv M(0)$ and $c$ is a dimensionless constant
which vanishes for an isotropic material. 

If we work to order $\epsilon$, the equations of
motion are linearized and we find that sinusoidal waves of  
the form $\zeta = 2a \cos (kx - \omega_0 t)$ propagate
over the surface.  $\omega_0 = \omega_0(k)$ is given by 
the linear dispersion relation
\be
\omega_0 = -\beta E_0 {{k^2}\over \sigma},
\label{linear_dispersion}
\ee
where $\beta\equiv qM_0\nu\Omega$ and $\sigma\equiv\tanh kh_0$ \cite{Krug1}.
The corresponding group velocity is
\be
v_g = v_0\left[h_0^2 k^2\left ( 1-{1\over{\sigma^2}}\right ) 
      +2{h_0k\over\sigma}\right ],
\label{group_velocity}
\ee
where $v_0 = -\beta E_0/h_0$ is the phase velocity
of waves of infinitely long wavelength.  If the amplitude $2a$ is not
a constant but instead varies slowly with position,
the resulting amplitude modulation will propagate with the group velocity.

The goal of this paper is to go beyond 
the linear approximation and study wave trains of finite amplitude. 
To begin, we will find the analog of the Stokes wave propagating
over deep water: we will consider the 
limit $h_0\to \infty$ and restrict our attention to
periodic wave trains with constant amplitude $a$.  These simplifying assumptions will be relaxed in the next section. 

Expanding $\zeta$ and $\phi$ in powers of the amplitude,
we have
\be
\zeta = [a {\cal E} + \mu_2 a^2 {\cal E}^2 + \mu_3 a^3 {\cal E}^3+\ldots]
	+{\rm c.c.}
\label{zeta_exp}
\ee
and
\be
\phi = [\nu_1 a e^{kz} {\cal E} 
+ \nu_2 a^2 e^{2kz} {\cal E}^2 
+ \nu_3 a^3 e^{3kz} {\cal E}^3
+ \tilde\nu_3 a^3 e^{kz} {\cal E}
+\ldots]
+{\rm c.c.},
\label{phi_exp}
\ee
where ${\cal E}\equiv e^{i(kx -\omega t)}$,
\lq\lq c.c." denotes the complex conjugate, and 
$\mu_2$, $\mu_3$, $\nu_1$, $\nu_2$, $\nu_3$ and $\tilde\nu_3$
are constants to be determined.  We can take $a$ to be
real without loss of generality.   
$\omega$ must depend on $a$ if the appearance
of secular terms at the third order is to be avoided. 
Again expanding in powers of $a$, we have
\be
\omega = \omega_0 + a^2 \omega_2 +\ldots,
\label{omega_exp}
\ee
where $\omega_0 = -\beta E_0 k^2$ is the linear dispersion
relation for $h_0\to\infty$ and $\omega_2 = \omega_2(k)$.

Our expansion for $\phi$ satisfies the Laplace equation.
Since we are seeking a periodic wave train, the boundary conditions
(\ref{orig_loc_h}) and (\ref{orig_loc_phi}) do not apply.
Inserting our expansions (\ref{zeta_exp}), (\ref{phi_exp})
and (\ref{omega_exp}) into Eqs. (\ref{orig_topbc}) and (\ref{orig_eom}),
setting $\alpha = 1- 2c$,
and working to third order in $a$, we obtain
\be
\omega_2 = \beta E_0 \left (1+ {3\over 2}\alpha - \alpha^2 \right) k^4,
\ee
\be
\mu_2 = {1\over 2}\alpha k,
\ee
\be
\mu_3 = {1\over 4}\alpha (2\alpha - 1 ) k^2,
\ee
\be
\nu_1=-i E_0,
\ee
\be
\nu_2= i\left( 1- {1\over 2}\alpha \right ) E_0 k,
\ee
\be
\nu_3 = -{1\over 2} i  
\left ( 3 - {{7\over 2}}\alpha + \alpha^2 \right ) E_0 k^2,
\ee
and
\be
\tilde\nu_3 = {1\over 2} i (3\alpha -1) E_0 k^2.
\ee

The nonlinear dispersion relation is
\be
\omega = -\beta E_0 k^2\left [ 1 - \left (1+ {3\over 2}\alpha - \alpha^2 \right) 
k^2 a^2 +\ldots \right ].
\label{thick_film_dispersion}
\ee
For the isotropic case $\alpha = 1$, the magnitude of the phase velocity 
\be
v_p \equiv {\omega \over k} = -\beta E_0 k \left ( 1 - {3\over 2} k^2 a^2 +\ldots \right )
\ee
is a {\it decreasing} function of $a$ when $a$ is small.  In contrast,
the phase velocity of small amplitude gravity waves on deep water
increases with amplitude.  For a given amplitude $2a$, the 
phase velocity is smallest for $\alpha = 3/4$, or, equivalently,
for $c=1/8$.

The form of the periodic wave train is given by
\be
\zeta = 2a \cos(kx-\omega t) + \alpha ka^2 \cos 2(kx-\omega t)
+{1\over 2}\alpha (2\alpha -1) k^2 a^3 \cos 3(kx-\omega t)
+\ldots
\label{stokes}
\ee
To third order in $a$, the wave is symmetric about the crests,
even though the applied electric field breaks the right-left symmetry.

{\bf IV. Slow Modulation of Finite Amplitude Periodic Wave Trains}

We will now relax the simplifying assumptions of Section III.
Suppose $h_0$ is finite and that the amplitude is a slowly
varying function of position.  We set
\be
\zeta = \sum_{n=-\infty}^\infty \zeta_n E^n
\label{zeta_expansion}
\ee
and
\be
\phi = \sum_{n=-\infty}^\infty \phi_n E^n,
\label{phi_expansion}
\ee
where $E\equiv e^{i(kx-\omega_0 t)}$
and $\omega_0$ is given by the linear dispersion relation
for films of finite thickness, Eq. (\ref{linear_dispersion}).  
Since $\zeta$ and $\phi$ are
real, $\zeta_{-n}=\zeta_n^\ast$ and $\phi_{-n}=\phi_n^\ast$ for all $n$.

Because the amplitude of the wave is of order $\epsilon$, the coefficients
$\zeta_n$ and $\phi_n$ are of order $\epsilon^n$.  We may therefore write
\be
\zeta_n=\sum_{j=n}^\infty \epsilon^j \zeta_{nj}
\ee
and 
\be
\phi_n=\sum_{j=n}^\infty \epsilon^j \phi_{nj}.
\ee
Here $n\ge 0$ and the $\zeta_{nj}$'s and $\phi_{nj}$'s do not depend on $\epsilon$.
Note as well that $\zeta_{00}=\phi_{00}=0$ and that the $\zeta_{0j}$'s and $\phi_{0j}$'s
are all real.

To find a steady, periodic wave train (the analog of the Stokes wave),
we would take the amplitudes $\zeta_{nj}$ to be constants and the $\phi_{nj}$'s to depend
on $z$ alone.  We will go further, and allow the $\zeta_{nj}$'s to vary slowly in space
and time as viewed from the frame of reference moving with the group velocity.
As a result, we will be able to find the steady, periodic
wave train and investigate its stability as well.  
(Our analysis closely parallels Davey and Stewartson's treatment of 
modulated gravity waves \cite{DS}.)  To be explicit, we set
\be
\zeta_{nj}=\zeta_{nj}(\xi,\tau)
\ee
and 
\be
\phi_{nj}=\phi_{nj}(\xi, z, \tau),
\ee
where $\xi\equiv\epsilon(x-v_g t)$ and $\tau\equiv\epsilon^2 t$.  The method
of multiple scales will be employed, i.e., $x$, $\xi$, $t$, and $\tau$ will be
treated as independent variables.  Using the chain rule, we see that
\be
{d\over{dx}} = {\partial\over{\partial x}}+\epsilon{\partial\over{\partial\xi}}
\label{partial_x_exp}
\ee
and
\be
{d\over{dt}}= {\partial\over{\partial t}}-\epsilon v_g {\partial\over{\partial\xi}}
			+\epsilon^2 {\partial\over{\partial\tau}}.
\label{partial_t_exp}
\ee

We now insert our expansions for $d/dx$ and $\phi$ into the Laplace equation (\ref{orig_Laplace}). 
This results in a series of ordinary differential equations for the $\phi_{nj}$'s.  Applying
the boundary condition (\ref{orig_bottombc}), we find that
\be
\phi_{11}=A{{\cosh kz'}\over{\cosh kh_0}},
\ee
\be
\phi_{22}=F{{\cosh 2kz'}\over{\cosh 2kh_0}},
\ee
\be
\phi_{12}=G{{\cosh kz'}\over{\cosh kh_0}} - iA_\xi {{z'\sinh kz'}\over{\cosh kh_0}},
\ee
and
\be
\phi_{13} = -{1\over 2} A_{\xi\xi} (z')^2 {{\cosh kz'}\over{\cosh kh_0}}
		-iG_\xi z' {{\sinh kz'}\over{\cosh kh_0}} + H {{\cosh kz'}\over{\cosh kh_0}},
\ee
where $A$, $F$, $G$ and $H$ depend only on $\xi$ and $\tau$.  It also follows that
$\phi_{01}$ and $\phi_{02}$ are functions of $\xi$ and $\tau$ alone, while
\be
{{\partial\phi_{03}}\over {\partial z}}= -z' {{\partial^2\phi_{01}}\over {\partial\xi^2}}.
\ee

The next step is quite laborious but is nonetheless straightforward, 
and so I will omit the details.
Inserting our expansions for $d/dx$, $d/dt$, $\zeta$ and $\phi$ into
Eqs. (\ref{orig_topbc}) and (\ref{orig_eom}) and working to second order
in $\epsilon$, we find that
\be
\zeta_{01}=0,
\ee
\be
\zeta_{11}=i{{\sigma}\over{E_0}} A,
\ee
\be
\zeta_{12}=i {{\sigma}\over{E_0}} G + {{h_0}\over{E_0}} A_\xi,
\label{zeta12}
\ee
\be
\zeta_{22}=-{k\over{\sigma E_0^2}}\left [1-\left (1-{1\over 2}\alpha\right )\sigma^2\right ]A^2
\ee
and
\be
F={{ik}\over{2E_0\sigma^2}}(1+\sigma^2)\left [1-\left (2-{1\over 2}\alpha\right )\sigma^2\right ]A^2.
\ee

Equating the coefficient of $\epsilon^3 E^0$ to zero in Eqs. (\ref{orig_topbc}) and (\ref{orig_eom}) yields two linear equations for $\zeta_{02}$ and 
$\partial\phi_{01}/\partial\xi$.  Solving these for $\zeta_{02}$, we obtain
\be
\zeta_{02}=\left ({{\sigma k}\over {E_0^2}}\right )
{{\alpha\sigma h_0 k -2}\over{v_g/v_0-1}} \vert A\vert^2.
\ee

When we equate the coefficient of $\epsilon^3 E^1$ to zero in Eqs. (\ref{orig_topbc}) and (\ref{orig_eom}), we obtain two equations that are consistent only if
\be
i A_\tau = \Gamma A_{\xi\xi} + \Lambda \vert A\vert^2 A,
\label{nls}
\ee
where the constants $\Gamma$ and $\Lambda$ are given by
\be
\Gamma= - {1\over 2} {{d^2\omega_0}\over {dk^2}}
={{\beta E_0}\over\sigma}\left (1+2\sigma h_0 k - {{v_g}\over {v_0}}\right )
\label{mu}
\ee
and
\be
\Lambda  
={{\beta k^4}\over{\sigma E_0}}\left [
{2\over{\sigma^2}} - 1 -\alpha + \left ({5\over 2}\alpha - \alpha^2\right)\sigma^2
-\left (1-\sigma^2+{\sigma\over{h_0 k}} \right)
{{2v_g/v_0 - \alpha\sigma h_0 k }\over{v_g/v_0-1}}\right ].
\ee

Eq. (\ref{nls}) is the nonlinear Schr\"odinger equation, and describes
the time development of the amplitude modulation.  It has nonlinear
plane wave solutions
\be
A=A_0 e^{i(\Omega\tau - \kappa\xi )},
\label{plane_wave}
\ee
where $A_0$ and $\kappa$ are constants and 
$\Omega = \Gamma \kappa^2 - \Lambda\vert A_0\vert^2$.  Let us recast
the solution with $\kappa=0$ in terms of the original, physical variables.
For $\kappa=0$, Eq. (\ref{zeta12}) has the solution $\zeta_{12}=G=0$,
and we will specialize to this case.  We set $\sigma \epsilon A_0 = -iE_0 a$,
where $a$ is a real constant of order $\epsilon$.  To second order
in $a$, 
\be
\zeta = 2a\cos(kx-\omega t)
	+\left ( {k\over\sigma} \right ){{\alpha\sigma h_0 k -2}\over{v_g/v_0-1}} a^2
	+{{2k}\over{\sigma^3}} \left [1-\left (1-{1\over 2}\alpha\right )\sigma^2\right]
	a^2 \cos 2(kx-\omega t),
\label{stokes2}
\ee
where
\be
\omega = \omega_0 + \Lambda \left ({{E_0}\over\sigma}\right )^2 a^2.
\label{dispersion}
\ee
Eq. (\ref{stokes2}) describes a periodic wave train of finite amplitude propagating over a metal or semiconducting layer of finite thickness.  
Eq. (\ref{dispersion}) is the nonlinear dispersion relation.

In the limit $h_0 k\gg 1$, Eq. (\ref{stokes2}) becomes
\be
\zeta = 2a \cos(kx-\omega t) +  \alpha ka^2 \cos 2(kx-\omega t) +{1\over 2} \alpha k a^2 +\ldots
\label{stokes3}
\ee
and Eq. (\ref{dispersion}) reduces to Eq. (\ref{thick_film_dispersion}).  Eq. (\ref{stokes3}) reduces
to Eq. (\ref{stokes}) after a trivial vertical shift of the origin, and so the results of this
section have the correct thick film limit.

For $h_0 k\ll 1$, Eq. (\ref{stokes2}) is
\be
\zeta = 2a \cos(kx-\omega t) -{2\over {h_0^3 k^2}} a^2 \left [1 - \cos 2(kx-\omega t) \right ] +\ldots
\label{stokes4}
\ee
where
\be
\omega = v_0 k\left ( 1+{1\over 3}h_0^2 k^2 + {2\over {h_0^4 k^2}} a^2 +\ldots \right ).
\label{thin_film_dispersion}
\ee
Eqs. (\ref{stokes4}) and (\ref{thin_film_dispersion}) are valid for $ka\ll (kh_0)^3\ll 1$.
It has been shown elsewhere \cite {soliton} 
that the equation of motion for small amplitude, long
waves is the Korteweg-de Vries (KdV) equation
\be
\zeta_t = - v_0 \zeta_x + {1\over 3}v_0 h_0^2 \zeta_{xxx} + 2 {v_0\over{h_0}} \zeta \zeta_x .
\label{KdV}
\ee
The KdV equation has finite amplitude, periodic solutions (cnoidal waves)
of the form
\be 
\zeta = -2a+4a \,{\rm cn}^2 [\pi^{-1} K(m)(kx-\omega t)\vert m ],
\label{cnoidal_wave}
\ee
where cn is a Jacobian elliptic function, $K(m)$ is the complete elliptic integral 
of the first kind,
\be
\omega = v_0 k  \left [ 1+ {{4a}\over {3h_0}} \left ({2\over m} -1\right )\right ],
\label{cnoidal_omega}
\ee
and $m$ is given implicitly by the following relation:
\be
m K^2(m) = {{2\pi^2 a}\over {h_0^3 k^2}}.
\label{relation}
\ee
Eqs. (\ref{cnoidal_wave}) - (\ref{relation}) give 
a good approximate solution to the original equations
of motion in the limit that $a/h_0$ and $h_0 k$ tend to zero with $a/h_0^3 k^2$ remaining
finite as the limit is taken.  If the amplitude of the cnoidal wave is small enough that
$ka\ll (kh_0)^3\ll 1$, Eqs. (\ref{cnoidal_wave}) and (\ref{cnoidal_omega}) reduce to
Eqs. (\ref{stokes4}) and (\ref{thin_film_dispersion}).  Thus, in the thin film limit, 
the nonlinear wave (\ref{stokes2}) is a cnoidal wave of small amplitude.

\begin{figure}
\includegraphics[scale=0.5]{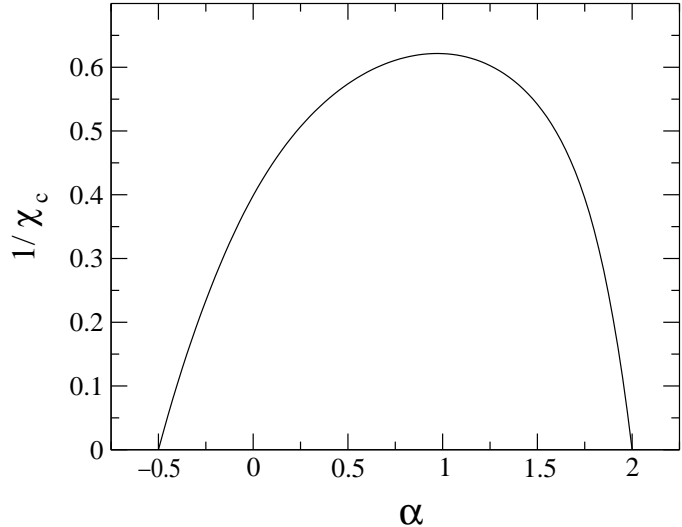}%
\caption{\label{Fig. 2} A plot of $1/\chi_c$ as a function of $\alpha$.  
$\chi_c^{-1}$ is zero for $\alpha\ge 2$ and $\alpha\le -1/2$.}
\end{figure}

The periodic wave train (\ref{plane_wave}) is marginally stable if $\Gamma\Lambda \le 0$ but
is unstable if $\Gamma\Lambda > 0$ \cite{HO}. 
This means that for a given value of $\alpha$, the periodic wave
train is unstable if $kh_0 $ exceeds a critical value we shall denote by $\chi_c=\chi_c(\alpha)$, 
and is marginally stable for $kh_0\le\chi_c$.  The critical value $\chi_c = 1.609927\ldots$ 
for the isotropic case ($\alpha = 1$) and is infinite for $\alpha \ge 2$ and $\alpha \le -1/2$.
Fig. \ref{Fig. 2} is a plot of $1/\chi_c(\alpha)$ as a function of $\alpha$.  It is
intereresting to note that the smallest value of $\chi_c$ is obtained
for a value of $\alpha$ differing just slightly from unity
($\alpha = 0.972\ldots$).

For $k h_0 \ll 1$,
the product $\Gamma\Lambda$ is negative, regardless of the value of $\alpha$.  The periodic
wave train is therefore marginally stable in this limit.  This is in accord with the
fact that cnoidal waves are marginally stable \cite{Drazin}.

{\bf IV. Conclusions}

In this paper, the electromigration-induced dynamics of small-amplitude disturbances on a low index crystal surface were studied. 
Periodic wave trains of finite amplitude were found, as well
as their dispersion relation.
These wave trains are unstable for $kh_0 > \chi_c(\alpha)$, 
and are otherwise marginally stable.  The value of the parameter
$\alpha$ depends on strength of the material anisotropy, and is 
equal to 1 if the anisotropy is negligible.  For $\alpha =1$,
$\chi_c = 1.609927\ldots$.  For $\alpha\le -1/2$ and $\alpha\ge 2$,
on the other hand, $\chi_c$ is infinite and
the wave train is marginally stable no matter what its wavelength.

The equation of motion for slow modulations
of the finite amplitude, periodic wave train was shown to be 
the nonlinear Schr\"odinger equation.  
As is well known, this equation is exactly solvable for initial
conditions that vanish with sufficient speed as 
$\vert x\vert\to\infty$ \cite{ZS}.  This exact solution has 
an number of consequences for electromigration-induced propagation 
of surface waves.  If
$k h_0 > \chi_c(\alpha)$, a localized initial wave packet of arbitrary
shape will eventually disintegrate into a number of envelope
solitons and a small oscillatory tail.  These solitons survive collisions
with each other with no permanent change except a shift in phase and
position.

What are the prospects of an experimental test of the 
theory developed in this paper?
Quite recently, the SEM-induced
dynamics of metal surfaces of large area have
received some attention, but unfortunately these experiments
were carried out on polycrystalline films, and so they do not
permit a test of the theory \cite{Shimoni,Shimoni2}.

The experimental situation is more promising in the case of semiconductors.
A number of beautiful experimental studies of the SEM-induced dynamics
of single crystal silicon films have been carried out, but
on initially planar, {\it vicinal} surfaces \cite{Latyshev,bunching}.  
These studies revealed that
a vicinal surface is unstable against step-bunching
for one current direction, and that it is
stable for the opposite current direction. Experimental studies of
the dynamics of perturbed, low-index silicon surfaces that are subject to
high electric fields have not yet been
carried out.  

To test the theory, 
a singly-periodic periodic wave structure could be etched into 
a low-index silicon surface.  After perturbing
the surface in this way, a high electrical current 
would be applied parallel to the ripple wavevector.
The subsequent surface dynamics could be imaged by scanning tunneling
microscopy, as has already been done for vicinal surfaces.  

The analog of the Stokes wave would be obtained 
by etching many parallel ripples into the wafer surface
before the application of current.  In this way, the
amplitude dependence of the phase velocity could be
determined and compared with the theoretical prediction.  To observe the
development of an envelope soliton (or solitons), on the 
other hand, a few tens or hundreds of ripples
would be etched into the film initially.  This would produce a 
\lq\lq wave packet" that would ultimately disintegrate 
into one or more envelope solitons and a small oscillatory tail
under the action of an applied current.

\end{document}